\documentclass{article}

\usepackage[final]{neurips_2021}
\usepackage{amsmath}
\usepackage{natbib}
\usepackage[utf8]{inputenc} 
\usepackage[T1]{fontenc}    
\usepackage{hyperref}       
\usepackage{url}            
\usepackage{booktabs}       
\usepackage{amsfonts}       
\usepackage{nicefrac}       
\usepackage{microtype}      
\usepackage{xcolor}         

\usepackage{textcomp}
\usepackage{graphicx,color}
\usepackage{algorithm}
\usepackage{algorithmic}
\usepackage{bm}
\usepackage{soul}
\usepackage{colortbl}
\usepackage{multirow}
\usepackage{caption}
\usepackage{titlesec}

\usepackage[normalem]{ulem}
\definecolor{orange}{rgb}{1,0.5,0}
\definecolor{red}{rgb}{1, 0, 0}

\newcommand{\hisaxy}{\textbf{\textit{HiSAXy }}}
\title{HiSAXy: A fast methodology for solar wind structure identification in millions of time series}

\author{
  Hala Lamdouar \\
  University of Oxford \\
  \And
  Sairam Sundaresan \\
  Intel Labs\\
  \And
  Anna Jungbluth \\
  University of Oxford \\
  \And
  Sudeshna Boro Saikia \\
  University of Vienna \\
  \And
  Amanda Joy Camarata \\
  Colorado School of Mines \\
  \And
  Nathan Miles \\
  University of Colorado, Boulder \\
  \And 
  Marcella Scoczynski \\
  Federal University of Technology – Paran\'a \\
  \And
  Mavis Stone \\
  Stanford University \\
  \And
  Anthony Sarah \\
  Intel Labs\\
  \And
  Andr\'es Mu\~noz-Jaramillo \\
  Southwest Research Institute\\
  \And
  Ayris Narock \\
  NASA Goddard Space Flight Center \\
  ADNET Systems Inc \\
  \And
  Adam Szabo \\
  NASA Goddard Space Flight Center \\
}

\begin{document}

\maketitle

\begin{abstract}
We present a hybridized unsupervised clustering algorithm \hisaxy as a novel way to identify frequently occurring magnetic structures embedded in the interplanetary magnetic field (IMF) carried by the solar wind. The \hisaxy algorithm utilizes a combination of indexable Symbolic Aggregate approXimation (iSAX) and Hierarchical Density-Based Spatial Clustering of Applications with Noise (HDBSCAN) to efficiently identify clusters of patterns embedded in time series data. We utilized \hisaxy to identify small-scale structures, known as discontinuities, embedded in time series measurements of the IMF. In doing so, we demonstrate the capability of the algorithm to significantly reduce the amount of human analysis hours required to identify these structures, all the while maintaining a high degree of self similarity within a given cluster of time series data. 
\end{abstract}


\section{Introduction}
\label{introduction}



The solar wind is a constant stream of plasma the originates from the outermost regions of the Sun's atmosphere known as the solar corona. As the solar wind plasma expands in the solar system, it carries with it a remnant of the solar magnetic field that is formally referred to as the interplanetary magnetic field (IMF). The solar wind plasma itself is highly structured and as a result, so is the IMF. The structures embedded in the solar wind span large scales in both space and time; the largest structures last on the order of days to weeks, while the smallest structures last on the order of seconds to minutes. Previous applications of artificial intelligence to the solar wind have focused on classifying the solar wind into distinct sub-types (fast or slow) based on the plasma properties, e.g., solar wind speed and number density \cite{heidrichmeinser2018kmeans,amaya2020swclassification}. In this work, we use unsupervised learning techniques to identify and classify repeating magnetic structures in the solar wind.  This task is typically done by hand in heliophysics, significantly limiting the scope and breadth of the resulting classification.  Our goal is to find an optimal compromise between the specificity of our classification and how long it takes a human scientist to analyze and interpret the output of our algorithm.

Unsupervised supervised machine learning techniques can help classify inputs in the absence of known labels. This is accomplished by searching for clusters in a feature space comprised of the inputs.  In this project we use build on the Hierarchical density-based spatial clustering of applications with noise (HDBSCAN)\citep{mcinnes2017hdbscan}, which is an extension of DBSCAN\citep{ester1996density}. HDBSCAN works by finding kernels of density in phase space, which are gradually grown to encompass nearby points enabling the merging of adjacent density based clusters. This is accomplished by creating a hierarchical tree where small individual peaks of density gradually merge into larger clusters until all points have been aggregated together.

The novelty of our approach is to sort all time series segments using the indexable Symbolic Aggregate approXimation (iSAX) algorithm\cite{isaxTB2008}, prior to the application of HDBSCAN.  iSAX is a novel multi-resolution symbolic representation of time series that enables the scaling and indexing of massive datasets, with millions of time series, requiring a single operation per time-series, and enabling ultra-fast retrieval and pattern matching \cite{8955558}.  It is constructed by encoding time series segments into words of specified length and of a specified cardinally (SAX)\cite{isax2003}, and sorting these words into a binary search tree. This tree grows organically, increasing the cardinality of the SAX representation.  We demonstrate that, while an iSAX tree is a superb framework for pattern matching and retrieval, it is suboptimal for unsupervised pattern identification. Our hybridized approach, \textit{\textbf{HiSAXy}}, combines the strengths of iSAX (speed and scalability) and HDBSCAN (robust clustering) to produce a highly efficient unsupervised clustering algorithm that saves an incredible amount of human effort in pattern discovery and interpretation.


\begin{figure}[t]
    \centering
    \includegraphics[width=\textwidth]{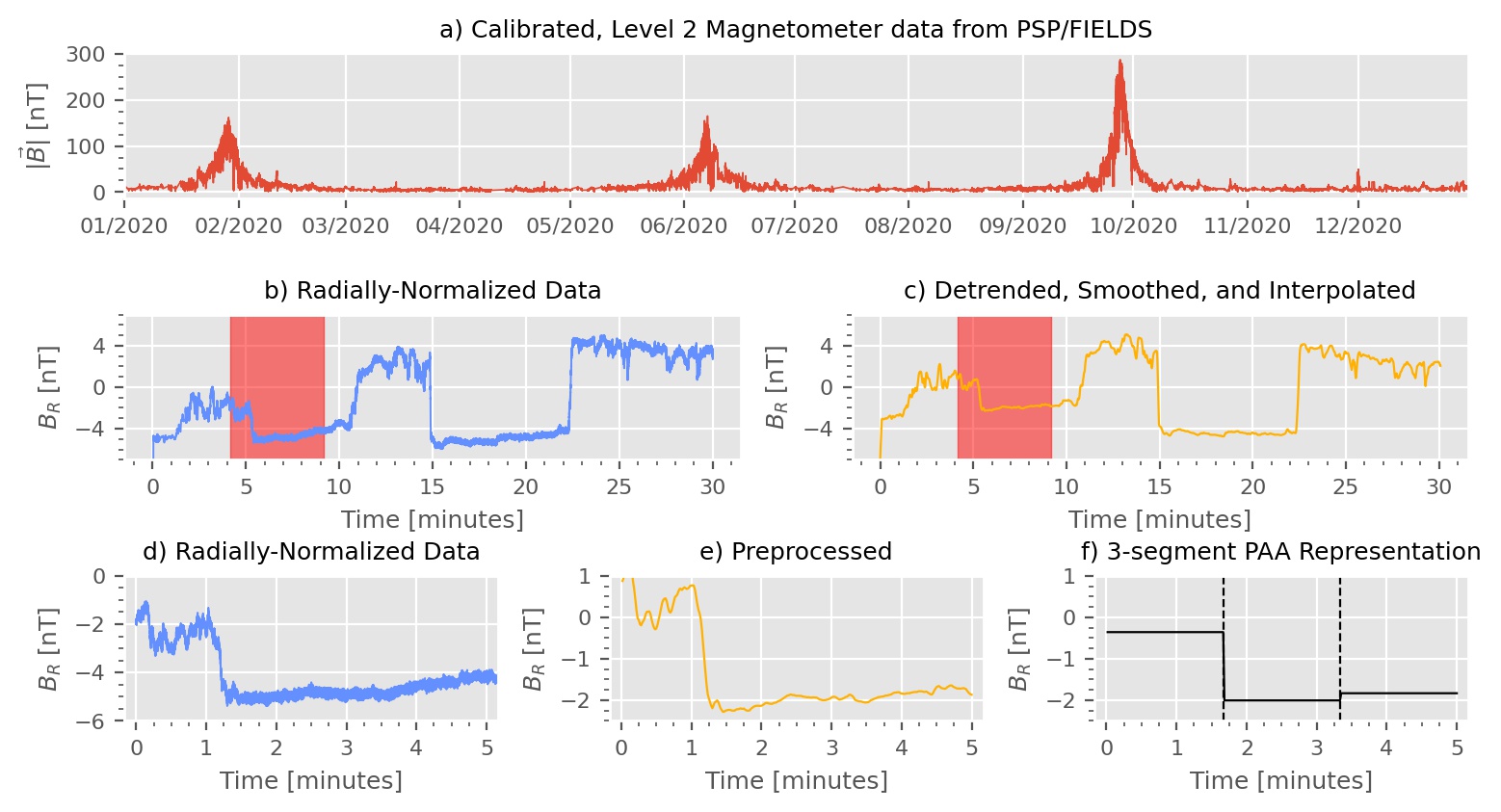}
    \caption{\textbf{a)} The original calibrated, level 2 magnetometer data from the PSP/FIELDS instrument for the year 2020. \textbf{b)} 30 minute interval of the magnetic field data after applying radial scaling. \textbf{c)} The same 30 minute interval after detrending using a rolling mean over an 1.5 hr interval, smoothing using a rolling mean over a 1 second interval, and linearly interpolating to a uniform cadence of 1 Hz. \textbf{d)} Zoom in to the 5 minute interval highlighted by the red shading in the top left. \textbf{e)} Zoom in to the 5 minute interval highlighted by the red shading in the top right. \textbf{f)} Piece-wise aggregate approximation of the 5 minute interval using three, 100 second intervals.}
    \label{fig:data_transformations}
\end{figure}

\section{Datasets}
\label{datasets}

We analyzed magnetometer data obtained by the FIELDS \cite{pspfields2016} instrument on board the Parker Solar Probe (PSP) from October 2, 2018 to December 31, 2020. This data can be downloaded through the Space Physics Data Facility’s Coordinated Data Analysis Website \footnote{\href{https://cdaweb.gsfc.nasa.gov/index.html/}{https://cdaweb.gsfc.nasa.gov/index.html/}}{} (CDAWeb). In total, we analyzed 150,488 5-minute segments of time series data containing $\sim$2.2 billion in-situ measurements of the three vector components of the IMF embedded in the solar wind. For each 5-minute segment, we apply the following preprocessing steps. First, we apply the appropriate radial-scaling laws to the components of the magnetic field to remove the effects of PSP's orbit on the baseline magnetic field strength evident in Fig.~\ref{fig:data_transformations}a). Next, we detrend the data with an 1.5hr window to remove any local trends on times scales that are 18 times longer than discontinuities. After that, we smooth with a rolling mean over a one-second window and linearly interpolate to a one-second cadence to create the cleaned times series. The end result of preprocessing can be seen in Figure \ref{fig:data_transformations}e.

\begin{figure}[t]
  \centering
  \includegraphics[width=\textwidth]{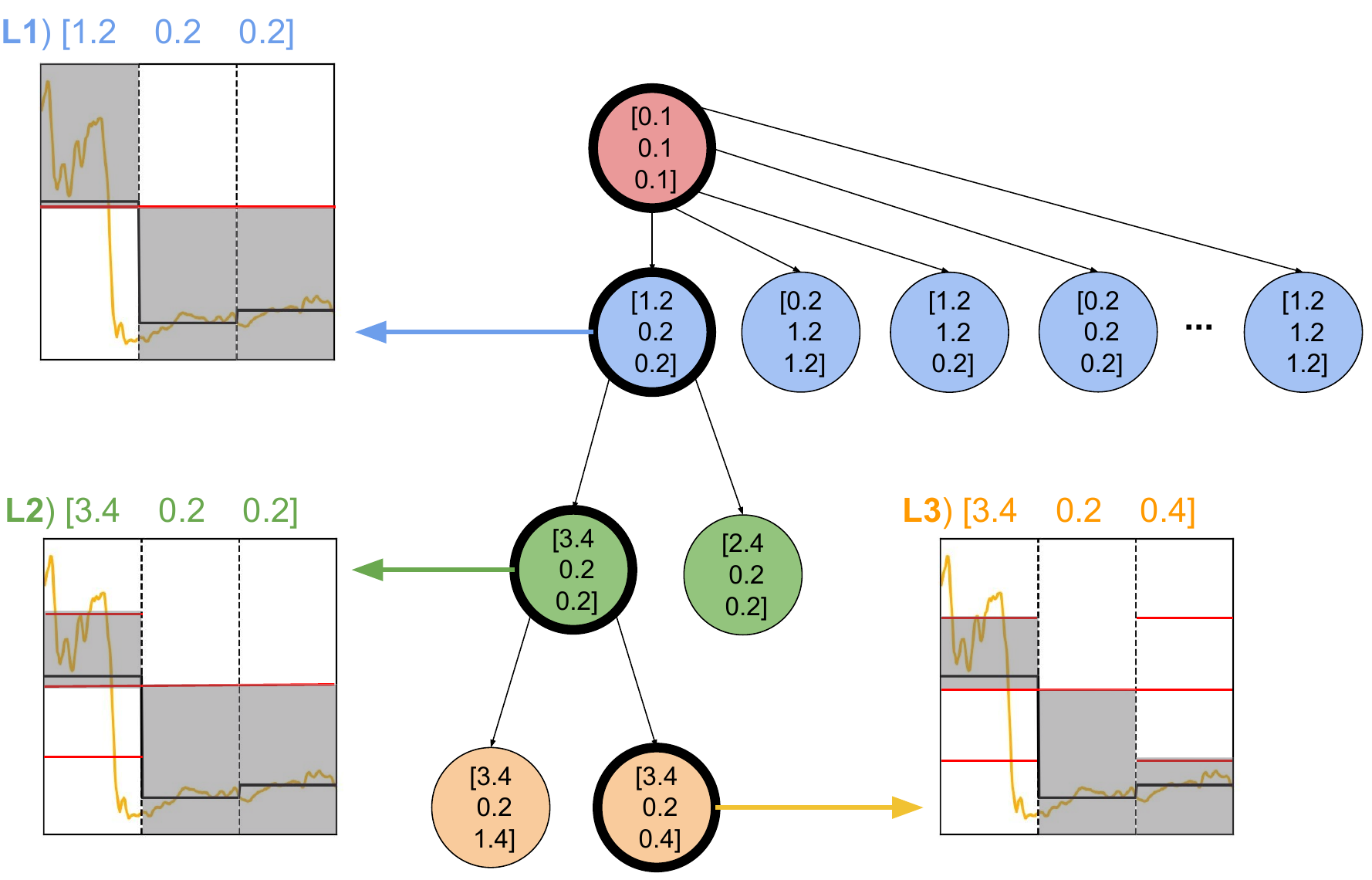}
  \caption{iSAX tree for indexing sequences using 3-letter words. Tree depth is represented using different colors. iSAX increases the depth of the tree by escalating the cardinality of one of the letters in each node-split. A black highlight denotes the tree node containing the sample time series at each tree level.  An iSAX letter is composed of two numbers separated by a dot: range (\textit{\textbf{r}}) and cardinality (\textit{\textbf{c}}) -- \textbf{\textit{r.c}}. The sample time series is represented by the iSAX word [1.2 0.2 0.2] (\textbf{L1}), [3.4 0.2 0.2] (\textbf{L2}), and [3.4 0.2 0.4] (\textbf{L3}). Shaded areas indicate the range of each letter for the sample time series in a given cardinality and can be use to approximate the euclidean distance between any two time series in the tree, enabling the clustering of tree nodes using HDBSCAN. The sample sequence is the same shown in Figs.~\ref{fig:data_transformations}-e \& f.  Tick marks and labels have been removed for simplicity.
    \label{fig:iSax-Tree}}
\end{figure}

\section{Methodology}
\label{method}



\hisaxy first indexes all the time series segments using iSAX.  iSAX creates a tree structure, illustrated in Figure \ref{fig:iSax-Tree}, with a root node (red), which contains pointers to children nodes that can be internal nodes (pointing to children of their own) or terminal nodes (holding onto indexed segments).  Each node is characterized by an iSAX word that encodes well defined ranges in magnetic fields and times (see shaded regions in Fig.~\ref{fig:iSax-Tree}).  The first tree level (blue in Fig.~\ref{fig:iSax-Tree}) contains nodes with iSAX words at a minimum cardinality of two. This amounts to initially sorting time series depending on whether each letter in their PAA representation is below or above the mean value. Each terminal node can hold a maximum number of segments determined by the user. Whenever that number is exceeded, terminal nodes spawn two children that each contain a portion of the segments and become internal nodes themselves. The new nodes inherit the iSAX word of their parent, with the exception of a single letter which is escalated in cardinality, enabling a more nuanced description of the segments it holds.  The example of Fig.~\ref{fig:iSax-Tree} shows how this escalation leads from the starting iSAX word with the minimum cardinality of 2 (L1: [1.2 0.2 0.2]) into an iSAX word with one letter of cardinality 4 at level 2 (L2 [3.4 0.2 0.2]), and two letters of cardinality 4 at level 3 (L3 [3.4 0.2 0.4]).  

\begin{figure}[thb]
  \centering
  \includegraphics[width=\textwidth]{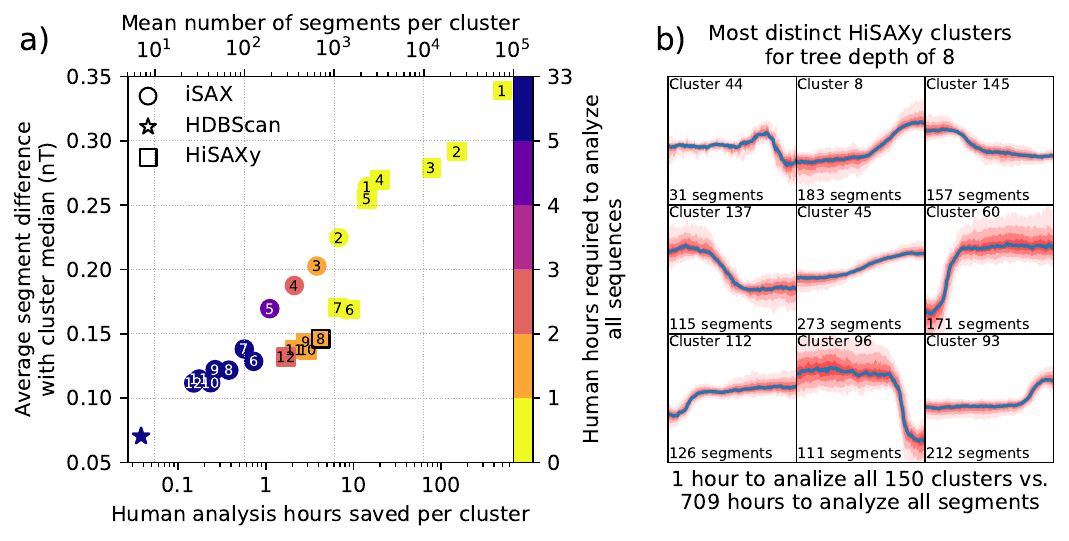}
  \caption{\textbf{a)} iSAX (circles) can be used as an approximate clustering algorithm by extracting all the nodes at a specified tree depth (numbers inside markers) and treat them as clusters.  A deeper (larger) tree level has more specific clusters with smaller average differences between members (bottom of y-axis). However, each cluster also has fewer segments so human analysis still requires significant effort (left of x-axis).  Coupling iSAX with HDBSCAN (HiSAXy; squares) combines similar nodes into clusters, reducing the number of clusters that need to be analyzed, without significantly impacting cluster specificity.  Resulting in significant saving of human effort.  HDBSCAN alone (star) produces even more clusters than iSAX alone, which is undesirable for our application. \textbf{b)} We find that combining iSAX and HDBSCAN for nodes of tree level 8 is a very good compromise between specificity and human effort.  This approaches saves more than 700 hours of human effort compared to working with the original segments.  \label{fig:isax_HDBScan_comp}}
\end{figure}

At any specified tree level, all time series segments indexed underneath a node are more similar to each other than those under any other node at the same level. However, tree escalation is dependent on the order in which the segments are provided and is strongly dependent on the user specified threshold.  This is problematic when using it as a clustering mechanism.   We address this by clustering all nodes at a specified level using HDBSCAN.  We do this by using the euclidean distance between nodes given the expected value of each letter in their iSAX word.

\hisaxy is built on the iSAX implementation of Lucas Foulon \cite{foulonThesis2020}, with the addition of a framework that enables the retrieval of the origin of each time series segment in the database.  For HDBSCAN, we use the open source python package HDBSCAN \cite{mcinnes2017hdbscan}.  Our codebase can be found in a github repository to be released after the anonymized review.   Our experiments were run on the Google Cloud Platform using the Compute Engine with a c2-standard-8 VM (8vCPUs, 32GB RAM) and 1 TB of SSD disk space.

\section{Experiments and Results}
\label{experiments}

In this work, we compare the results of clustering the time series data using iSAX, HDBSCAN, and \hisaxy using different tree level depths.  Focus on identifying an optimal combination of specificity and human effort that enables scientific discovery at an unprecedented scale in heliophysics. To asses cluster quality, we use a measure of self similarity defined as follows,
\begin{equation}
   \alpha = \frac{1}{T} \sum_{t=0}^{T} s(t) = \frac{1}{T} \sum_{t=0}^{T} \left( \frac{1}{N_{\textrm{s}}}\sum_{i=1}^{N_{\textrm{s}}} |B_{\textrm{med}}(t) - B_{i}(t)| \right),
\end{equation}
where $N_s$ is the number of segments in a cluster, $B_{i}(t)$ is the $i^{\mathrm{th}}$ segment of magnetic ftield time-series data, and $B_{\mathrm{med}}(t)$ is the median of all the segments at each time step, $t$. To assess human effort, we visually inspected \hisaxy clusters to label clusters containing magnetic discontinuities.  We registered an average rate of $\sim150$ clusters per hour per human. Using this classification rate, we asses the time saved in analyzing clusters of similar curves generated by the three algorithms. In Figure \ref{fig:isax_HDBScan_comp}, we show the results of the comparison which demonstrates the unique ability of \hisaxy to generate an optimal number of segments per cluster (thereby increasing the time saved) without sacrificing specificity. 








\section{Conclusions, Limitations, and Future Work}
\label{conclusions}
In this paper, we introduced a hybridized unsupervised learning approach, \hisaxy, for classifying structures embedded in the solar wind. When compared to iSAX and HDBSCAN individually, we found that \hisaxy outperforms both by identifying larger clusters without sacrificing the intracluster self similarity. At the present moment, the fundamental limitation of \hisaxy is the implicit assumption that the time series data follow a Gaussian distribution. However, this is not a hard constraint and can be modified as needed. In the future, we plan on extending our work by applying \hisaxy to study a variety solar wind properties across a range of timescales. 




\section*{Broader Impact}
\label{impact}
 
The ability to automate the identification of aperiodic structures embedded in the solar wind has been a long-standing goal for the heliophysics community. Current methodologies rely on heuristic techniques that can result in discrepancies in the identification of even the largest structures like ICMEs \cite{russell2005defining}. As the archive of solar wind data continues to grow with data from new missions like PSP and Solar Orbiter, so does the burden on helio- and space physicists. By utilizing the \hisaxy framework, researchers can focus their efforts on analyzing the most frequently occurring magnetic structures at arbitrary timescales (5-minute intervals, 30-minute intervals, 12-hour intervals, etc...). Furthermore, our work is applicable to not only time series measurements of other solar wind properties like number density, velocity, and temperature, but also any form of time series data.

\section*{Acknowledgments and Disclosure of Funding}
\label{acknowledgments}
This work was conducted at the Frontier Development Laboratory (FDL) USA 2021. The FDL USA is a public / private research partnership between NASA, the SETI Institute and private sector partners including Google Cloud, Intel, IBM, Lockheed Martin, and NVIDIA. These partners provide the data, expertise, training, and compute resources necessary for rapid experimentation and iteration in data-intensive areas.

\bibliography{main}

\end{document}